\documentclass[conference]{IEEEtran}
\ifCLASSINFOpdf
\else
\fi
\usepackage{url}

\usepackage{amsmath}
\usepackage{amssymb}

\usepackage{amsthm}

\theoremstyle{plain}
\newtheorem{theorem}{Theorem}
\newtheorem{lemma}[theorem]{Lemma}
\newtheorem{corollary}[theorem]{Corollary}

\newtheorem{definition}[theorem]{Definition}

\renewcommand{\labelenumi}{(\roman{enumi})}

\newcommand{\abs}[1]{\left\lvert#1\right\rvert}

\DeclareMathOperator{\Dom}{dom}

\DeclareMathOperator{\graph}{Graph}

\newcommand{\N}{\mathbb{N}}
\newcommand{\Z}{\mathbb{Z}}
\newcommand{\Q}{\mathbb{Q}}
\newcommand{\R}{\mathbb{R}}
\newcommand{\X}{\{0,1\}^*}

\begin{document}
%
\title{
Properties of Optimal Prefix-Free Machines
as Instantaneous Codes}

\author{\IEEEauthorblockN{Kohtaro Tadaki}
\IEEEauthorblockA{Research and Development Initiative, Chuo University\\
1-13-27 Kasuga, Bunkyo-ku, Tokyo 112-8551, Japan\\
Email: tadaki@kc.chuo-u.ac.jp\quad
WWW: http://www2.odn.ne.jp/tadaki/}}


%


\maketitle

\begin{abstract}
\boldmath
The
optimal prefix-free machine $U$ is a universal decoding algorithm
used to define the notion of program-size complexity $H(s)$
for a finite binary string $s$.
Since the set
of all halting inputs for $U$ is chosen to form a prefix-free set,
the optimal prefix-free machine
$U$
can be regarded
as an instantaneous code for noiseless source coding scheme.
In this paper,
we investigate the properties of optimal prefix-free machines
as instantaneous codes.
In particular,
we investigate the properties of the set $U^{-1}(s)$ of
codewords associated with a symbol $s$.
Namely,
we investigate the number of codewords in $U^{-1}(s)$
and the distribution of codewords in $U^{-1}(s)$
for each symbol $s$,
using the toolkit of algorithmic information theory.
\end{abstract}


%
\IEEEpeerreviewmaketitle

\section{Introduction}

Algorithmic information theory (AIT, for short) is a framework
for applying
information-theoretic and probabilistic ideas to recursive function theory.
One of the primary concepts of AIT is the \textit{program-size complexity}
(or \textit{Kolmogorov complexity}) $H(s)$ of a finite binary string $s$,
which is defined as the length of the shortest binary
input
for a universal decoding algorithm $U$,
called an \textit{optimal prefix-free machine},
to output $s$.
By the definition,
$H(s)$
can be thought of as
the information content of the individual finite binary string $s$.
In fact,
AIT has precisely the formal properties of
normal
information theory (see Chaitin~\cite{C75}).
On the other hand, $H(s)$ can also be thought to represent
the amount of randomness contained in a finite binary string $s$,
which cannot be captured
in a computational manner.
In particular,
the notion of program-size complexity plays a crucial role in
characterizing the \textit{randomness} of an infinite binary string,
or equivalently, a real.

The optimal prefix-free machine $U$ is chosen so as to satisfy that
the set $\Dom U$ of all halting inputs for $U$ forms a prefix-free set.
Therefore, as considered in Chaitin \cite{C75},
we can think
of the optimal prefix-free machine $U$ as a decoding equipment at the receiving end of
a noiseless binary communication channel.
We can regard
its programs
(i.e.,
finite binary strings in $\Dom U$)
as codewords and
can regard the result of the computation by $U$,
which is a finite binary string,
as a decoded ``symbol.''
Since $\Dom U$ is a prefix-free set,
such codewords form what is called an ``instantaneous code,''
so that successive symbols sent through the channel
in the form of concatenation of codewords can be separated.%
\footnote{Note that AIT does not assume the existence
of an encoding algorithm $E$ such that
$E(s)=p$ if and only if $U(p)=s$.}

Thus, from the point of view of information theory,
it is important to investigate
the properties of optimal prefix-free machine as an instantaneous code.
In this paper, in particular,
we investigate the properties of the set $U^{-1}(s)$ of
codewords associated with a symbol $s$,
where
$U^{-1}(s)=\{\,p\mid U(p)=s\,\}$.
Unlike for instantaneous codes in normal information theory,
the codeword $p$ associated with each symbol $s$ by $s=U(p)$
is not necessarily unique for optimal prefix-free machines $U$ in AIT.
We investigate this property from various aspects.

After the preliminary section,
in Section~\ref{nocw}
we investigate the number of codewords in $U^{-1}(s)$.
We show the following:
(i) While keeping $H(s)$ unchanged for all $s$, we can modify $U$
so that each $U^{-1}(s)$ is a finite set,
where the number of codewords in $U^{-1}(s)$ is
bounded to the above by some total recursive function $f(s)$,
i.e., by some computable function $f(s)$. 
(ii) This upper bound $f(s)$ cannot be chosen to be tight at all.
(iii) As a result,
even in the case where all $U^{-1}(s)$ are a finite set,
the number of codewords in $U^{-1}(s)$ is not bounded to the above
on all finite binary strings $s$.
(iv) While keeping $H(s)$ unchanged for all $s$, we can modify $U$
so that each $U^{-1}(s)$ is an infinite set.
In Section~\ref{docw},
we then 
investigate the distribution of codewords in $U^{-1}(s)$.
We estimate the distribution
using the
notion of
program-size complexity,
and then show that the estimation
is
tight.

%


\section{Preliminaries}
\label{preliminaries}

\subsection{Basic Notation}
\label{basic notation}

We start with some notation about numbers and strings
which will be used in this paper.
$\#S$ is the cardinality of $S$ for any set $S$.
$\N=\left\{0,1,2,3,\dotsc\right\}$ is the set of natural numbers,
and $\N^+$ is the set of positive integers.
$\Q$ is the set of rationals, and
$\R$ is the set of reals.
%
Normally,
$O(1)$ denotes any function $f\colon \N^+\to\R$ such that
there is $C\in\R$ with the property that
$\abs{f(n)}\le C$ for all $n\in\N^+$.

$\X=
\left\{
  \lambda,0,1,00,01,10,11,000,\dotsc
\right\}$
is the set of finite binary strings
where $\lambda$ denotes the \textit{empty string},
and $\X$ is ordered as indicated.
We identify any string in $\X$ with a natural number in this order,
i.e.,
we consider $\varphi\colon \X\to\N$ such that $\varphi(s)=1s-1$
where the concatenation $1s$ of strings $1$ and $s$ is regarded
as a dyadic integer,
and then we identify $s$ with $\varphi(s)$.
For any $s \in \X$, $\abs{s}$ is the \textit{length} of $s$.
A subset $S$ of $\X$ is called
\textit{prefix-free}
if no string in $S$ is a prefix of another string in $S$.
For any function $f$,
the domain of definition of $f$ is denoted by $\Dom f$.
We write ``r.e.'' instead of ``recursively enumerable.''



\subsection{Algorithmic Information Theory}
\label{ait}

In the following
we concisely review some definitions and results of
AIT
\cite{C75,C87b,N09,DH10}.
A \textit{prefix-free machine} is a partial recursive function
$C\colon \X\to \X$
such that
$\Dom C$ is a prefix-free set.
For each prefix-free machine $C$ and each $s \in \X$,
$H_C(s)$ is defined by
\begin{equation*}
  H_C(s) =
  \min
  \left\{\,
    \abs{p}\,\big|\;p \in \X\>\&\>C(p)=s
  \,\right\}
  \;\;
  \text{(may be $\infty$)}.
\end{equation*}
A prefix-free machine $U$ is said to be \textit{optimal} if
for each prefix-free machine $C$ there exists $d\in\N$
with the following property;
if $p\in\Dom C$, then there is $q$ for which
$U(q)=C(p)$ and $\abs{q}\le\abs{p}+d$.
Note that a prefix-free machine $U$ is optimal if and only if
for each prefix-free machine $C$ there exists $d\in\N$ such that,
for every $s\in\X$,
$H_U(s)\le H_C(s)+d$.
It is easy to see that there exists an optimal prefix-free machine.
We choose a particular optimal prefix-free machine $U$
as the standard one for use,
and define $H(s)$ as $H_U(s)$,
which is referred to as
the \textit{program-size complexity} of $s$ or
the \textit{Kolmogorov complexity} of $s$.
It follows that
for every prefix-free machine $C$ there exists $d\in\N$ such that,
for every $s\in\X$,
\begin{equation}\label{minimal}
  H(s)\le H_C(s)+d.
\end{equation}
Based on this we can show that,
for every partial recursive function $\Psi\colon \X\to \X$,
there exists $d\in\N$ such that,
for every $s \in \Dom \Psi$,
\begin{equation}\label{Psi}
  H(\Psi(s))\le H(s)+d.
\end{equation}
Based on \eqref{minimal}
we can also show that there exists $c\in\N$ such that,
for every $n\in\N^+$,
\begin{equation}\label{eq: fas2}
  H(n)\le 2\log_2 n+c.
\end{equation}

For any $s\in\X$,
we define $s^*$ as $\min\{\,p\in\X\mid U(p)=s\}$,
i.e., the first element in the ordered set $\X$
of all strings $p$ such that $U(p)=s$.
Then, $\abs{s^*}=H(s)$ for every $s\in\X$.
For any $s,t\in\X$,
we define $H(s,t)$ as $H(b(s,t))$,
where $b\colon \X\times \X\to \X$ is
a particular bijective total recursive function.

AIT has precisely the formal properties of
normal
information theory, as demonstrated by Chaitin~\cite{C75}.
The program-size complexity $H(s)$ corresponds to
the notion of entropy in information theory,
while $H(s,t)$ corresponds to the notion of joint entropy in information theory.

The program-size complexity $H(s)$ is originally defined
using the notion of program-size, as in the above.
However,
it is possible to define $H(s)$ without referring to such a notion.
Namely,
as in the following,
we first introduce a \textit{universal probability} $m$,
and then define $H(s)$ as $-\log_2 m(s)$.
%
%
A universal probability is defined as follows.

\begin{definition}[universal probability, Zvonkin and Levin \cite{ZL70}]
A function $r\colon \X\to[0,1]$ is called
a \textit{lower-computable semi-measure} if
$\sum_{s\in \X}r(s)\le 1$ and
the set $\{(a,s)\in\Q\times\X\mid a<r(s)\}$ is r.e.
We say that a lower-computable semi-measure $m$ is
a \textit{universal probability} if
for every lower-computable semi-measure $r$,
there exists $c\in\N^+$ such that,
for all $s\in \X$, $r(s)\le cm(s)$.
\hfill\IEEEQED
\end{definition}

The following theorem can be then shown
(see e.g.~Chaitin \cite[Theorem 3.4]{C75} for its proof).

\begin{theorem}%
\label{eup}
For every optimal prefix-free machine $V$,
the function $2^{-H_V(s)}$ of $s$ is a universal probability.
\hfill\IEEEQED
\end{theorem}

For each universal probability $m$,
by Theorem~\ref{eup}
we see that $H(s)=-\log_2 m(s)+O(1)$ for all $s\in\X$.
Thus it is possible to define $H(s)$ as $-\log_2 m(s)$
with a particular universal probability $m$
instead of as $H_U(s)$.
Note that
the difference up to an additive constant is
nonessential
to AIT.

Normally, for each prefix-free machine $C$ and each $s\in\X$,
the set $C^{-1}(s)$ is defined by
\begin{equation*}
  C^{-1}(s)=\{\,p\in\Dom C \mid C(p)=s\,\}.
\end{equation*}
Note that $V^{-1}(s)\neq\emptyset$
for every optimal prefix-free machine $V$ and every $s\in\X$.

\section{The Number of Codewords}
\label{nocw}

In this section,
we investigate the properties of the number $\#V^{-1}(s)$ of codewords in $V^{-1}(s)$
for an optimal prefix-free machine $V$.
In Theorem~\ref{VWfinite} below we show that,
while keeping $H_V(s)$ unchanged for all $s$,
we can modify $V$ so that each $V^{-1}(s)$ is a finite set,
where $\#V^{-1}(s)$ is
bounded to the above by some total recursive function $f(s)$. 
Before that,
we prove a more general theorem for
prefix-free machines
in general,
as follows.

\begin{theorem}\label{CDfinite}
For every prefix-free machine $C$,
there exists a prefix-free machine $D$
for which the following conditions (i), (ii), and (iii) hold:
\begin{enumerate}
  \item $H_D(s)=H_C(s)$ for every $s\in\X$.
  \item $D^{-1}(s)$ is a finite set for every $s\in\X$.
  \item Moreover, there exists a partial recursive function
    $f\colon \X\to \N^+$ such that
    $\#D^{-1}(s)\le f(s)$ for every $s\in\Dom f$ and
    $\Dom f=\{\,s\in\X\mid D^{-1}(s)\neq\emptyset\,\}$.
\end{enumerate}
\end{theorem}

\begin{IEEEproof}
Let $C$ be an arbitrary prefix-free machine.
We define the \textit{graph} $\graph(C)$ of $C$ by
\begin{equation*}
  \graph(C)=\{\,(p,s)\in\X\times\X\mid C(p)=s\,\}.
\end{equation*}
Note that
$\graph(C)$ is an r.e.~set,
since $C\colon \X\to \X$ is a partial recursive function.
In the case where $\graph(C)$ is a finite set,
the set $\{\,s\in\X\mid C^{-1}(s)\neq\emptyset\,\}$ is finite
and the set $C^{-1}(s)$ is finite for every $s\in\X$.
Thus,
in this case,
by setting $D=C$ and $f(s)=\#C^{-1}(s)$,
the conditions (i), (ii), and (iii) hold,
and therefore the result follows.
Hence, in what follows
we assume that $\graph(C)$ is an infinite set.

Let $(p_1,s_1),(p_2,s_2),(p_3,s_3),\dotsc$ be
a particular recursive enumeration of the infinite r.e.~set $\graph(C)$.
It is then easy to show that
there exists a partial recursive function $g\colon \X\to \N^+$
which satisfies the following two conditions:
\renewcommand{\labelenumi}{(\alph{enumi})}
\begin{enumerate}
  \item $\Dom g=\{\,s\mid\exists\,i\in\N^+\;s_i=s\,\}$.
  \item $g(s) = \min\{\,i\in\N^+\mid s_i=s\,\}$ for every $s\in\Dom g$.
\end{enumerate}
\renewcommand{\labelenumi}{(\roman{enumi})}
We then define a partial recursive function $D\colon \X\to \X$
by the condition that
\begin{equation*}
  D^{-1}(s)
  =\left\{\,p_i\bigm| i\in\N^+\;\&\;s_i=s\;\&\;\abs{p_i}\le |p_{g(s)}|\,\right\}
\end{equation*}
if $s\in\Dom g$ and $D^{-1}(s)=\emptyset$ otherwise.
It is easy to see that such a partial recursive function $D$ exists.
By counting the number of binary strings of length at most $|p_{g(s)}|$,
we see that,
for each $s\in\Dom g$,
$\#D^{-1}(s)\le 2^{|p_{g(s)}|+1}-1$
and therefore $D^{-1}(s)$ is a finite set.
Thus, the condition (ii) holds for $D$.
Moreover,
by defining a partial recursive function $f\colon \X\to \N^+$
by the conditions that
$\Dom f=\Dom g$ and
$f(s)=2^{|p_{g(s)}|+1}-1$ for every $s\in\Dom f$,
the condition (iii) holds for $D$.

Next, we show that $D$ is a prefix-free machine.
It follows from the definition of $D$ that
\begin{equation}\label{D-1subsetC-1}
  D^{-1}(s)\subset C^{-1}(s)
\end{equation}
for every $s\in\X$.
Therefore we see that
\begin{equation*}
  \Dom D=\bigcup_{s\in\X} D^{-1}(s)
  \subset
  \bigcup_{s\in\X} C^{-1}(s)=\Dom C.
\end{equation*}
Thus, since $\Dom C$ is prefix-free,
its subset $\Dom D$ is also prefix-free.
Hence $D$ is a prefix-free machine.

Finally, we show that the condition (i) holds for $D$.
Let us assume that $C(p)=s$ and $\abs{p}=H_C(s)$.
Then $(p,s)\in\graph(C)$ and therefore $s\in\Dom g$.
Since $C(p_{g(s)})=s$,
we see that $\abs{p}\le |p_{g(s)}|$ and therefore $p\in D^{-1}(s)$.
Hence, $D(p)=s$ and therefore $H_D(s)\le\abs{p}$.
Thus we have
\begin{equation}\label{HDleHC}
  H_D(s)\le H_C(s)
\end{equation}
for every $s\in\X$.
On the other hand,
\eqref{D-1subsetC-1} implies that
\begin{equation}\label{HDgeHC}
  H_D(s)\ge H_C(s)
\end{equation}
for every $s\in\X$.
It follows from \eqref{HDleHC} and \eqref{HDgeHC} that
the condition (i) holds for $D$.
\end{IEEEproof}

\begin{theorem}\label{VWfinite}
For every optimal prefix-free machine $V$,
there exists an optimal prefix-free machine $W$
for which the following conditions (i), (ii), and (iii) hold:
\begin{enumerate}
  \item $H_W(s)=H_V(s)$ for every $s\in\X$.
  \item $W^{-1}(s)$ is a finite set for every $s\in\X$.
  \item Moreover, there exists a total recursive function
    $f\colon \X\to \N^+$ such that
    $\#W^{-1}(s)\le f(s)$ for every $s\in\X$.
\end{enumerate}
\end{theorem}

\begin{IEEEproof}
Let $V$ be an arbitrary optimal prefix-free machine.
Then it follows from Theorem~\ref{CDfinite} that
there exists a prefix-free machine $W$
for which the following conditions (a), (b), and (c) hold:
\renewcommand{\labelenumi}{(\alph{enumi})}
\begin{enumerate}
  \item $H_W(s)=H_V(s)$ for every $s\in\X$.
  \item $W^{-1}(s)$ is a finite set for every $s\in\X$.
  \item Moreover, there exists a partial recursive function
    $f\colon \X\to \N^+$ such that
    $\#W^{-1}(s)\le f(s)$ for every $s\in\Dom f$ and
    $\Dom f=\{\,s\in\X\mid W^{-1}(s)\neq\emptyset\,\}$.
\end{enumerate}
\renewcommand{\labelenumi}{(\roman{enumi})}
Therefore, the conditions~(i) and~(ii) hold obviously.
Since $V$ is optimal,
$W$ is also optimal by the above condition~(a).
On the other hand,
since $W$ is optimal,
$W^{-1}(s)\neq\emptyset$ for every $s\in\X$.
Thus, the condition (iii) holds.
\end{IEEEproof}

Through Theorems~\ref{main1} and \ref{main3} below,
we show that
the upper bound $f(s)$ in Theorem~\ref{VWfinite} cannot be chosen to be tight at all.
We first show a weaker result, Theorem~\ref{main1}.
Then, based on this,
we show a stronger result, Theorem~\ref{main3}.
The underlying idea of the proofs of Theorems~\ref{main1} and \ref{main3}
is due to A. R. Meyer and D. W. Loveland \cite[pp.~525--526]{L69}
(see also Chaitin~\cite[Theorem~5.1~(f)]{C75}).
In order to prove Theorem~\ref{main1},
we need Lemma~\ref{unbounded} below.
It is a well-known fact and
follows from the inequality
$\#\{\,s\in\X\mid H(s)<n\}\le 2^n-1$.

\begin{lemma}\label{unbounded}
Let $R$ be an infinite subset of $\X$.
Then the function $H(s)$ of $s\in R$ is not bounded to the above.
\hfill\IEEEQED
\end{lemma}


A function $f\colon\X\to\N$ is called \textit{right-computable} if
the set $\{\,(s,n)\in\X\times\N\mid f(s)\le n\,\}$ is r.e.
Obviously,
every total recursive function $f\colon\X\to\N$ is right-computable.

\begin{theorem}\label{main1}
Let $V$ be an optimal prefix-free machine,
and let $f\colon \X\to \N$.
Suppose that $\#V^{-1}(s)\le f(s)$ for all $s\in\X$ and $f$ is right-computable.
Then $\#V^{-1}(s)<f(s)$ for all but finitely many $s\in\X$.
\end{theorem}

\begin{IEEEproof}
We define a function $h$
by the following two conditions:
\renewcommand{\labelenumi}{(\alph{enumi})}
\begin{enumerate}
  \item $\Dom h=\{\,s\in\X\mid\#V^{-1}(s)=f(s)\,\}$.
  \item $h(s)=\min\{\,\abs{p}\mid p\in V^{-1}(s),\}$ for every $s\in\Dom h$.
\end{enumerate}
\renewcommand{\labelenumi}{(\roman{enumi})}
Note first that $V^{-1}(s)\neq\emptyset$ for every $s\in\X$
since $V$ is optimal.
Therefore $\min\{\,\abs{p}\mid p\in V^{-1}(s),\}$ is
well-defined as a natural number for every $s\in\X$.
Since $\#V^{-1}(s)\le f(s)$ for all $s\in\X$
and $f$ is right-computable,
it is easy to see that
the above two conditions (a) and (b) define
a partial recursive function $h\colon \X\to \N$.
On the other hand,
it follows from the
condition (b)
that
\begin{equation}\label{h=H}
  h(s)=H(s)
\end{equation}
for every $s\in\Dom h$.

Now, let us assume contrarily that
$\#V^{-1}(s)=f(s)$ for infinitely many $s\in\X$.
Then, obviously, $\Dom h$ is an infinite set.
It follows from Lemma~\ref{unbounded},
the function $h$ is not bounded to the above.
Thus,
given $n\in\N^+$,
by enumerating the graph of the partial recursive function $h$,
one can find $s\in\Dom h$ such that $n\le h(s)$.

Hence, combined with \eqref{h=H},
we see that
there exists a partial recursive function $\Psi\colon\N^+\to\X$
such that
$n\le H(\Psi(n))$.
Using \eqref{Psi},
we then see that $n\le H(n)+O(1)$ for all $n\in\N^+$.
It follows from \eqref{eq: fas2} that
$n\le 2\log_2 n+O(1)$ for all $n\in\N^+$.
Dividing by $n$ and letting $n\to\infty$ we have $1\le 0$, a contradiction.
This completes the proof.
\end{IEEEproof}

\begin{theorem}\label{main3}
Let $V$ be an optimal prefix-free machine,
and let $f\colon \X\to \N$.
Suppose that $\#V^{-1}(s)\le f(s)$ for all $s\in\X$ and $f$ is right-computable.
Then
\begin{equation*}
  \lim_{s\to\infty}\left\{f(s)-\#V^{-1}(s)\right\}=\infty.
\end{equation*}
Recall here that we identify $\X$ with $\N$.
\end{theorem}

\begin{IEEEproof}
We denote by $Q$ the set of all $k\in\Z$ such that
$k\le f(s)-\#V^{-1}(s)$
for all but finitely many $s\in\X$.
Note that $0\in Q$ and therefore $Q\neq\emptyset$.
This is because $\#V^{-1}(s)\le f(s)$ for all $s\in\X$.

Now, let us assume contrarily that
$f(s)-\#V^{-1}(s)$ does not diverge to $\infty$ as $s\to\infty$.
Then there exists $M\in\N$ such that, for infinitely many $s\in\X$,
$f(s)-\#V^{-1}(s)\le M$.
It is then easy to see that $k\le M$ for all $k\in Q$.
Thus,
since $Q$ is a nonempty subset of $\Z$ bounded to the above,
$Q$ has the maximum element $k_0$.
Since $k_0\in Q$,
\begin{equation}\label{k0fsV-1abfm}
  k_0\le f(s)-\#V^{-1}(s)
\end{equation}
for all but finitely many $s\in\X$.
If $k_0< f(s)-\#V^{-1}(s)$ for all but finitely many $s\in\X$,
then $k_0+1\in Q$ and
this contradicts the fact that $k_0$ is the maximum element of $Q$.
Thus, $k_0\ge f(s)-\#V^{-1}(s)$ for infinitely many $s\in\X$.
Hence, it follows from \eqref{k0fsV-1abfm} that
there exists a finite subset $E$ of $\X$ such that
$k_0\le f(s)-\#V^{-1}(s)$ for all $s\in\X\setminus E$
and 
$k_0= f(s)-\#V^{-1}(s)$ for infinitely many $s\in\X\setminus E$.

We define a function $g\colon \X\to \N$ by
$g(s)=\#V^{-1}(s)$ if $s\in E$
and $g(s)=f(s)-k_0$ otherwise.
Then, obviously,
$\#V^{-1}(s)\le g(s)$ for all $s\in\X$
and $g$ is right-computable.
Moreover,
$\#V^{-1}(s)=g(s)$ for infinitely many $s\in\X$.
However, this contradicts Theorem~\ref{main1},
and the proof is completed.
\end{IEEEproof}

\begin{corollary}\label{main2}
Let $V$ be an optimal prefix-free machine.
Suppose that $V^{-1}(s)$ is a finite set for all $s\in\X$.
Then the function $\#V^{-1}(s)$ of $s\in\X$ is not bounded to the above.
\end{corollary}

\begin{IEEEproof}
Assume contrarily that
the function $\#V^{-1}(s)$ of $s\in\X$ is bounded to the above.
Then there exists $M\in\N$ such that, for every $s\in\X$,
$\#V^{-1}(s)\le M$.
We define a function $f\colon \X\to \N$ by $f(s)=M$.
Then, obviously, $\#V^{-1}(s)\le f(s)$ for all $s\in\X$
and $f$ is right-computable.
It follows from Theorem~\ref{main3} that
$\lim_{s\to\infty}\left\{f(s)-\#V^{-1}(s)\right\}=\infty$.
However,
this contradicts the fact that
$f(s)-\#V^{-1}(s)\le M$ for all $s\in\X$.
This completes the proof.
\end{IEEEproof}

\begin{theorem}
For every optimal prefix-free machine $V$,
there exists an optimal prefix-free machine $W$
for which the following conditions (i) and (ii) hold:
\begin{enumerate}
  \item $H_W(s)=H_V(s)$ for every $s\in\X$.
  \item $W^{-1}(s)$ is an infinite set for every $s\in\X$.
\end{enumerate}
\end{theorem}

\begin{IEEEproof}
Let $V$ be an arbitrary optimal prefix-free machine.
We first show that
$V^{-1}(s_0)$ has at least two elements for some $s_0\in\X$.
In the case where
$V^{-1}(s_0)$ is an infinite set for some $s_0\in\X$,
obviously $V^{-1}(s_0)$ has at least two elements.
Thus, we assume that $V^{-1}(s_0)$ is a finite set for all $s_0\in\X$,
in what follows.

First,
it follows from Corollary~\ref{main2} that
$\#V^{-1}(s_0)\ge 2$ for some $s_0\in\X$.
Thus,
some $V^{-1}(s_0)$ has two elements $q$ and $r$ with $\abs{q}\ge \abs{r}$.
Let $b\colon \X\times \N\to \N$ be
a particular bijective total recursive function.
We then define a partial recursive function $W\colon \X\to \X$
by
the condition that
$W^{-1}(s)
=\left(V^{-1}(s)\setminus \{q\}\right)\cup
\{\,q0^{b(s,i)}1\mid i\in\N\,\}$
if $s=s_0$ and
$W^{-1}(s)=V^{-1}(s)\cup T(s)$
otherwise,
where $T(s)=\{\,q0^{b(s,i)}1\mid i\in\N\;\&\;H_V(s)\le\abs{q}+b(s,i)+1\,\}$.
Since the set $\{\,(s,n)\in\X\times\N\mid H_V(s)\le n\,\}$ is r.e.,
it is easy to see that such a partial recursive function $W$ exists.

Since $b$ is a bijection,
the set $\{\,q0^{b(s_0,i)}1\mid i\in\N\,\}$ is infinite and
the set
$T(s)$
is infinite for every $s\neq s_0$.
Therefore the condition (ii) holds for $W$.
On the other hand,
it follows that
\begin{equation}\label{WsubsetV-qq01}
\begin{split}
  &\Dom W
  =\bigcup_{s\in\X} W^{-1}(s)\\
  &\subset\Bigl(\Bigl(\bigcup_{s\in\X} V^{-1}(s)\Bigr)\setminus \{q\}\Bigr)
  \cup\{\,q0^{k}1\mid k\in\N\,\}\\
  &=\left(\Dom V\setminus \{q\}\right)\cup\{\,q0^{k}1\mid k\in\N\,\}.
\end{split}
\end{equation}
Thus, since $\Dom V$ is prefix-free and $q\in\Dom V$,
the most right-hand side of \eqref{WsubsetV-qq01} is prefix-free.
Hence its subset $\Dom W$ is also prefix-free,
and therefore $W$ is a prefix-free machine.

Finally,
we show that the condition (i) holds for $W$.
In the case of $s=s_0$,
since $\abs{q}<\abs{q0^k1}$ for all $k\in\N$ and
there is $r\in\Dom V$ with $\abs{r}\le \abs{q}$,
we have $H_W(s)=H_V(s)$.
In the case of $s\neq s_0$,
since the set
$T(s)$
does not contain any string of length less than $H_V(s)$,
we have $H_W(s)=H_V(s)$ again.
Thus, the condition (i) holds for $W$.
\end{IEEEproof}

\section{The Distribution of Codewords}
\label{docw}

In this section,
we investigate the distribution of codewords in $V^{-1}(s)$
for each optimal prefix-free machine $V$ and each $s\in\X$.
Solovay~\cite{Sol75} showed the following result for
the distribution of all codewords $\Dom V$
for an optimal prefix-free machine $V$.

\begin{theorem}\label{Solovay}
Let $V$ be an optimal prefix-free machine.
Then
\begin{equation*}
  \#\{\,p\in\X\mid \abs{p}\le n\;\&\;p\in\Dom V\,\}=2^{n-H(n)+O(1)}.
\end{equation*}
Namely,
there exists $d\in\N$ such that
\begin{enumerate}
  \item $\#\{\,p\in\X\mid \abs{p}\le n\;\&\;p\in\Dom V\,\}\le 2^{n-H(n)+d}$
    for all $n\in\N$, and
  \item $2^{n-H(n)-d}\le\#\{\,p\in\X\mid \abs{p}\le n\;\&\;p\in\Dom V\,\}$
    for all $n\in\N$ with $n-H(n)\ge d$.
    \hfill\IEEEQED
\end{enumerate}
\end{theorem}

Note that $\lim_{n\to\infty} \left\{n-H(n)\right\}=\infty$ by \eqref{eq: fas2}.
We refine
Theorem~\ref{Solovay}
to a certain extent.
For that purpose,
we define
\begin{equation*}
  S_C(n,s)=\{\,p\in\X\mid \abs{p}\le n\;\&\;C(p)=s\,\}
\end{equation*}
for each prefix-free machine $C$, each $n\in\N$, and each $s\in\X$.
We can then show the following theorem.

\begin{theorem}\label{maind1}
Let $C$ be a prefix-free machine.
Then
$\#S_C(n,s)\le 2^{n-H(n,s)+O(1)}$.
\end{theorem}

\begin{IEEEproof}
We show that there exists $d\in\N$ such that
$\#S_C(n,s)\le 2^{n-H(n,s)+d}$ for all $n\in\N$ and all $s\in\X$.
For that purpose,
we define a function $f\colon \N\to [0,\infty)$
by $f(b(n,s))=\#S_C(n,s)2^{-n-1}$.
Recall here that $b\colon \X\times \X\to \X$ is
a particular bijective total recursive function.
It is easy to see that
the set $\{(a,k)\in\Q\times\N\mid a<f(k)\}$ is r.e.
On the other hand,
\begin{equation*}
  \sum_{k=0}^{\infty}f(k)=\sum_{s\in\X}\sum_{n=0}^{\infty}\#S_C(n,s)2^{-n-1}
\end{equation*}
\begin{align*}
  &=\sum_{s\in\X}\sum_{n=0}^{\infty}\sum_{l=0}^n\#\overline{S_C}(l,s)2^{-n-1}\\
  &=\sum_{s\in\X}\sum_{l=0}^{\infty}\sum_{n=l}^{\infty}\#\overline{S_C}(l,s)2^{-n-1}\\
  &=\sum_{s\in\X}\sum_{l=0}^{\infty}\#\overline{S_C}(l,s)2^{-l}
  =\sum_{p\in\Dom C} 2^{-\abs{p}}\le 1,
\end{align*}
where $\overline{S_C}(n,s)=\{\,p\in\X\mid \abs{p}=n\;\&\;C(p)=s\,\}$.
Thus, $f$ is a lower-computable semi-measure.
It follows from Theorem~\ref{eup} that
there exists $d'\in\N$ such that $f(k)\le 2^{d'}2^{-H(k)}$ for all $k\in\N$.
Therefore we have
$\#S_C(n,s)2^{-n-1}\le 2^{d'-H(b(n,s))}$ for all $n\in\N$ and all $s\in\X$,
which implies that
$\#S_C(n,s)\le 2^{n-H(n,s)+d'+1}$ for all $n\in\N$ and all $s\in\X$,
as desired.
\end{IEEEproof}

%

Theorem~\ref{maind2} below shows that
the upper bound $2^{n-H(n,s)+O(1)}$ in Theorem~\ref{maind1}
is tight
among all optimal prefix-free machines.
In order to prove Theorems~\ref{maind2} and \ref{maind3} below,
we need the following lemma.

\begin{lemma}\label{Hs+n-Hs+ns=infty}
$H(s)+n-H(H(s)+n,s)$ diverges to $\infty$ as $n\to\infty$
uniformly on $s\in\X$.
Namely,
for every $M\in\N$, there exists $n_0\in\N$ such that,
for every $n\ge n_0$ and every $s\in\X$,
$H(s)+n-H(H(s)+n,s)\ge M$.
\end{lemma}

\begin{IEEEproof}
Let us consider a prefix-free machine $C$ such that,
for every $p,q\in\Dom U$, $C(pq)=b(\abs{p}+U(q),U(p))$,
where $U(q)$ is regarded as a natural number
based on our identification of $\X$ with $\N$.
It is easy to see that such a prefix-free machine exists.
For each $s\in\X$ and each $n\in\N$,
we see that $C(s^*n^*)=b(H(s)+n,s)$ and therefore
$H_C(b(H(s)+n,s))\le \abs{s^*n^*}=H(s)+H(n)$.
It follows from \eqref{minimal} that
there exists $d\in\N$ such that,
for every $s\in\X$ and every $n\in\N$,
$H(b(H(s)+n,s))\le H(s)+H(n)+d$.
Using \eqref{eq: fas2} we then see that
there exists $d'\in\N$ such that,
for every $s\in\X$ and every $n\in\N^+$,
$H(s)+n-H(H(s)+n,s)\ge n-2\log_2 n-d'$.
Hence, the result follows.
\end{IEEEproof}

\begin{theorem}\label{maind2}
There exists an optimal prefix-free machine $V$ which
satisfies that
$\#S_V(n,s)=2^{n-H(n,s)+O(1)}$.
Namely,
there exist an optimal prefix-free machine $V$ and $d\in\N$ such that
\begin{enumerate}
  \item $\#S_V(n,s)\le 2^{n-H(n,s)+d}$ for all $n\in\N$ and all $s\in\X$, and
  \item $2^{n-H(n,s)-d}\le \#S_V(n,s)$
    for all $n\in\N$ and all $s\in\X$ with $n-H(n,s)\ge d$.
\end{enumerate}
\end{theorem}

\begin{IEEEproof}
By Theorem~\ref{maind1}, it is enough to show that
the condition~(ii) holds for some optimal prefix-free machine $V$ and some $d\in\N$
(in fact, $d$ can be chosen to be $0$ in the following construction of $V$).

Let us consider a partial recursive function $V\colon \X\to \X$
such that, for every $p,s\in\X$,
$V(p)=s$ if and only if there exist $q,t\in\X$
for which $p=qt$ and $U(q)=b(\abs{p},s)$.
Since $U$ is a prefix-free machine and
$b\colon \X\times \X\to \X$ is a bijective total recursive function,
it is easy to see that
such a partial recursive function $V\colon \X\to \X$ exists.
Since $\Dom U$ is prefix-free and $b$ is an injective function,
we can also check that $\Dom V$ is prefix-free.
Thus $V$ is a prefix-free machine.

We show that $2^{n-H(n,s)}\le \#S_V(n,s)$
for all $n\in\N$ and all $s\in\X$ with $n-H(n,s)\ge 0$.
For each $n\in\N$ and $t\in\X$, if $\abs{t}=n-H(n,s)$,
then $\abs{b(n,s)^* t}=n$ and $V(b(n,s)^* t)=s$.
Recall here that $\abs{b(n,s)^*}=H(n,s)$.
Thus, for each $n\in\N$, if $n-H(n,s)\ge 0$
then $2^{n-H(n,s)}\le \#S_V(n,s)$,
as desired.

Finally, we show that $V$ is optimal.
By Lemma~\ref{Hs+n-Hs+ns=infty},
we see that there exists $n_0\in\N$ such that,
for every $s\in\X$, $H(s)+n_0-H(H(s)+n_0,s)\ge 0$.
Hence, for each $s\in\X$,
$\abs{b(H(s)+n_0,s)^* t}=H(s)+n_0$ and
therefore $V(b(H(s)+n_0,s)^* t)=s$,
where $t=0^{H(s)+n_0-H(H(s)+n_0,s)}$.
Thus, we see that $H_V(s)\le H(s)+n_0$ for all $s\in\X$,
which implies that $V$ is optimal.
This completes the proof.
\end{IEEEproof}

As
a complement
to Theorem~\ref{maind2},
the following theorem
shows that
only an optimal prefix-free machine can attain
the upper bound $2^{n-H(n,s)+O(1)}$ in Theorem~\ref{maind1}.

\begin{theorem}\label{maind3}
Let $C$ be a prefix-free machine.
Suppose that $2^{n-H(n,s)+O(1)}\le \#S_C(n,s)$,
namely, suppose that there exists $d\in\N$ such that
$2^{n-H(n,s)-d}\le \#S_C(n,s)$
for all $n\in\N$ and all $s\in\X$ with $n-H(n,s)\ge d$.
Then $C$ is optimal.
\end{theorem}

\begin{IEEEproof}
It follows from Lemma~\ref{Hs+n-Hs+ns=infty} that
there exists $n_0\in\N$ such that,
for every $s\in\X$, $H(s)+n_0-H(H(s)+n_0,s)\ge d$.
By the assumption, we see that, for each $s\in\X$,
$1\le 2^{H(s)+n_0-H(H(s)+n_0,s)-d}\le \#S_C(H(s)+n_0,s)$.
Thus, for each $s\in\X$,
$S_C(H(s)+n_0,s)\neq\emptyset$ and therefore there exists
$p\in\X$ such that $\abs{p}\le H(s)+n_0$ and $C(p)=s$.
Hence, we see that $H_C(s)\le H(s)+n_0$ for all $s\in\X$,
which implies that $C$ is optimal.
\end{IEEEproof}

%

\section*{Acknowledgments}
This work was supported
by KAKENHI, Grant-in-Aid for Scientific Research (C) (20540134),
by SCOPE
from the Ministry of Internal Affairs and Communications of Japan,
and by CREST from Japan Science and Technology Agency.



%

\end{document}